\documentclass[%
superscriptaddress,
floats,
floatfix,
twocolumn,
nofootinbib,
amsmath,amssymb,
aps,
]{revtex4-1}

\usepackage[left=1in, right=1in, top=1in, bottom=1.2in]{geometry}
\usepackage[usenames, dvipsnames]{color}
\usepackage[colorlinks, plainpages=false,citecolor=Green
]{hyperref}
\usepackage{graphicx}
\usepackage{xspace}
\usepackage{amssymb}
\usepackage[normalem]{ulem} 
\usepackage{fourier}
\usepackage{letltxmacro}
\usepackage{amsfonts}
\usepackage{amsmath}
\usepackage{amssymb}
\usepackage{bm}
\usepackage{dcolumn}
\usepackage{epsfig}
\usepackage{graphicx}
\usepackage{graphics}
\usepackage[latin1]{inputenc}
\usepackage{latexsym}
\usepackage{rotating}
\usepackage[usenames]{color}
\usepackage{float}
\usepackage{mathrsfs}
\usepackage{url}
\usepackage{times}

\def\Msun{M_\odot}

\def\prd{{\it Physical Review} D}

\def\apjl{{\it Astrophysical Journal} Letters}
\def\apj{{\it Astrophysical Journal}}

\def\lesssim{\mathrel{\hbox{\rlap{\hbox{\lower4pt\hbox{$\sim$}}}\hbox{$<$}}}}
\def\gtrsim{\mathrel{\hbox{\rlap{\hbox{\lower4pt\hbox{$\sim$}}}\hbox{$>$}}}}
\def\alt{\mathrel{\hbox{\rlap{\hbox{\lower4pt\hbox{$\sim$}}}\hbox{$<$}}}}
\def\agt{\mathrel{\hbox{\rlap{\hbox{\lower4pt\hbox{$\sim$}}}\hbox{$>$}}}}

\def\araa{{\it ARA\&A}}

\def\gta{\ifmmode {\mathbin{\lower 3pt\hbox   
			{$\,\rlap{\raise 5pt\hbox{$\char'076$}}\mathchar"7218\,$}}}
	\else {${\mathbin{\lower 3pt\hbox
				{$\rlap{\raise 5pt\hbox{$\char'076$}}\mathchar"7218\,$}}}
		$}\fi}
\def\lta{\ifmmode {\,\mathbin{\lower 3pt\hbox   
			{$\,\rlap{\raise 5pt\hbox{$\char'074$}}\mathchar"7218\,$}}}
	\else {${\mathbin{\lower 3pt\hbox
				{$\rlap{\raise 5pt\hbox{$\char'074$}}\mathchar"7218\,$}}}
		$}\fi}

\newcommand{\beq}{\begin{equation}}
\newcommand{\eeq}{\end{equation}}
\newcommand{\bea}{\begin{eqnarray}}
\newcommand{\eea}{\end{eqnarray}}

\newcommand{\NCSA}{\affiliation{NCSA, University of Illinois at Urbana-Champaign, Urbana, Illinois, 61801\vspace{.3in}}}
\newcommand{\ANCSA}{\affiliation{Department of Astronomy, University of Illinois at Urbana-Champaign, Urbana, Illinois, 61801}}

\begin{document}

	\title{{\normalfont }  {Deep Learning for Real-time Gravitational Wave Detection and Parameter Estimation: \\Results with Advanced LIGO Data}
	}
	
	
	\author{Daniel George}\ANCSA \NCSA 
	\author{E.~A. Huerta}\NCSA
	
	\begin{abstract}
		The recent Nobel-prize-winning detections of gravitational waves from merging black holes and the subsequent detection of the collision of two neutron stars in coincidence with electromagnetic observations have inaugurated a new era of multimessenger astrophysics. To enhance the scope of this emergent field of science, we pioneered the use of deep learning with convolutional neural networks, that take time-series inputs, for rapid detection and characterization of gravitational wave signals. This approach, \texttt{Deep Filtering}, was initially demonstrated using simulated LIGO noise. In this article, we present the extension of \texttt{Deep Filtering} using real data from LIGO, for both \textit{detection} and \textit{parameter estimation} of gravitational waves from binary black hole mergers using continuous data streams from multiple LIGO detectors. We demonstrate for the first time that machine learning can detect and estimate the true parameters of real events observed by LIGO. Our results show that \texttt{Deep Filtering} achieves similar sensitivities and lower errors compared to matched-filtering while being far more computationally efficient and more resilient to glitches, allowing real-time processing of weak time-series signals in non-stationary non-Gaussian noise with minimal resources, and also enables the detection of new classes of gravitational wave sources that may go unnoticed with existing detection algorithms. This unified framework for data analysis is ideally suited to enable coincident detection campaigns of gravitational waves and their multimessenger counterparts in real-time. \\

	\end{abstract}
	\maketitle

	
	
	\section{Introduction} 
	\label{intro}
	
	The first detection (GW150914) of gravitational waves (GWs), from the merger of two black holes (BHs), with the advanced Laser Interferometer Gravitational-wave Observatory (LIGO)~\cite{LSC:2015} has set in motion a scientific revolution~\cite{DI:2016} leading to the Nobel prize in Physics in 2017. This and subsequent groundbreaking discoveries~\cite{secondBBH:2016,bbhswithligo:2016,thirddetection,2017arXiv170909660T} were brought to fruition by a trans-disciplinary research program at the interface of experimental and theoretical physics, computer science and engineering as well as the exploitation of high-performance computing (HPC) for numerical relativity simulations~\cite{gr,NRI:2016,2016CQGra..33u5004U} and high-throughput computing facilities for data analysis~\cite{huerta:2017boss,2017arXiv170506202W}. 
	
	The recent detection of the binary black hole (BBH) merger (GW170814) with a three-detector network enabled new phenomenological tests of general relativity regarding the nature of GW polarizations, while significantly improving the sky localization of this GW transient~\cite{2017arXiv170909660T}. This enhanced capability to localize GW transients provided critical input for the first detection of GWs from the merger of two neutron stars (NSs) and in conjunction with follow-up observations across the electromagnetic (EM) spectrum~\cite{BNSdet:2017}. This multimessenger event has finally confirmed that NS mergers are the central engines of short gamma ray bursts~\cite{Eichler:1989,Paczynski:1986,Narayan:1992,Kochanek:1993mw,sum:2009CQGra,phi:2009astro2010S}. 
	
	Matched-filtering, the most sensitive GW detection algorithm used by LIGO, currently targets a 3D parameter space (compact binary sources with spin-aligned components on quasi-circular orbits)~\cite{2013PhRvD..87j4028G,Carl:2016arXiv,CR:2015PRL}---a subset of the 8D parameter space available to GW detectors~\cite{Anto:2015arXiv,Naoz:2013,Samsing:2014,Huerta:2017a,Lehner:2014a}. Recent studies also indicate that these searches may miss GWs generated by compact binary populations formed in dense stellar environments~\cite{Sergey:2016,Huerta:2017a,Huerta:2014,Huerta:2013a}. Extending these template-matching searches to target spin-precessing, quasi-circular or eccentric BBHs is computationally prohibitive~\cite{2016PhRvD..94b4012H}. 
	
	Based on the aforementioned considerations, we need a new paradigm to overcome the limitations and computational challenges of existing GW detection algorithms. An ideal candidate would be the rapidly advancing field called Deep Learning, which is a highly scalable machine learning technique that can learn directly from raw data, without any manual feature engineering, by using deep hierarchical layers of ``artificial neurons'', called neural networks, in combination with optimization techniques based on back-propagation and gradient descent~\cite{DL-Nature,DL-Book}. Deep learning, especially with the aid of GPU computing, has recently achieved immense success in both commercial applications and artificial intelligence (AI) research~\cite{DL-Nature,DL-Review,DNN-Medicine,AlphaGoZero,DeepStack,WaveNet,ConvNet-Radio}. 
	
	Our technique, \texttt{Deep Filtering}~\cite{DeepFiltering}, employs a system of two deep convolution neural networks (CNNs~\cite{lecun98-cnn}) that directly take time-series inputs for both classification and regression. In our foundational article~\cite{DeepFiltering}, we provided a comprehensive introduction to the fundamental concepts of deep learning and CNNs along with a detailed description of this method. Our previous results showed that CNNs can outperform traditional machine learning methods, reaching sensitivities comparable to matched-filtering for directly processing highly noisy time-series data streams to detect weak GW signals and estimate the parameters of their source in real-time, using GW signals injected into \textit{simulated} LIGO noise. 
	
	In this article, we extend \texttt{Deep Filtering} to analyze GW signals in \textit{real} LIGO noise.  We demonstrate, for the first time, that Deep Learning can be used for both signal detection and multiple-parameter estimation directly from extremely weak time-series signals embedded in highly non-Gaussian and non-stationary noise, once trained with some templates of the expected signals. Our results show that deep CNNs achieve performance comparable to matched-filtering methods, while being several orders of magnitude faster and far more resilient to transient noise artifacts such as glitches. We also illustrate how \texttt{Deep Filtering} can deal with data streams of arbitrary length from multiple detectors. Most importantly, this article shows for the very \textit{first} time that machine learning can successfully detect and recover true parameters of \textit{real} GW signals observed by LIGO. Furthermore, we show that after a single training process, \texttt{Deep Filtering} can automatically generalize to noise having new Power Spectral Densities (PSDs) from different LIGO events, without re-training.
	
	Our results indicate that \texttt{Deep Filtering} can interpolate between templates, \textit{generalize} to new classes of signals beyond the training data, and, surprisingly, detect GW signals and measure their parameters even when they are contaminated by glitches. We present experiments demonstrating the robustness of Deep Filtering in the presence of glitches, which indicate its applicability in the future for glitch classification and clustering efforts essential for GW detector characterization. Deep learning, in principle, can learn characteristics of noise in the LIGO detectors and develop better strategies than matched-filtering, which is known to be only optimal for Gaussian noise. Since all the intensive computation is diverted to the one-time training stage of the CNNs, template banks of practically any size may be used for training after which continuous data streams can be analyzed in real-time with a single CPU, while very intensive searches can be rapidly carried out using a single GPU. \texttt{Deep Filtering} can also be used to instantly narrow down the parameter space of GW detections, which can then be followed up with existing pipelines using a few templates around the predicted parameters, thus accelerating GW analysis with minimal computational resources across the full parameter space of signals. 
	
	
	\section{Methods}
	\label{meth}
	
	\texttt{Deep Filtering} consists of two steps, involving a classifier CNN and a predictor CNN, with similar architectures, as described in our previous article~\cite{DeepFiltering}. The classifier has an additional softmax layer which returns probabilities for True or False depending on whether a signal is present. The classifier is first applied to the continuous data stream via a sliding window. If the classifier returns higher probability for True, the predictor is applied to the same input to determine the parameters of the source. In a multi-detector scenario, the \texttt{Deep Filtering} CNNs may be applied separately to each data stream and the coincidence of detections with similar parameters would strengthen the confidence of a true detection, which can then be verified quickly by matched-filtering with the predicted templates.
	
	In this work, we have used injections of GW templates originating from quasi-circular, non-spinning, stellar-mass BBH systems, which LIGO/Virgo is expected to detect with the highest rate~\cite{bel:2016Na}. We assumed the source is optimally oriented with respect to the detectors which reduces our parameter-space to the two individual masses of the BBH system, which we restricted in the range $5\Msun$ to $75\Msun$ such that their mass-ratios were between 1 and 10. In the same manner as before, we fixed the input duration to 1 second, and a sampling rate of 8192Hz, which is more than sufficient for the events we are considering. These are arbitrary choices, as the input size of the CNNs can be easily modified to take inputs with any duration or sampling rate from any number of detectors.
	
	
	The datasets of waveform templates used to train and test our CNNs were obtained using the open-source, effective-one-body (EOB) code~\cite{Tara:2014}. Our training set contained about 2500 templates, with BBHs component masses sampled in the range $5\Msun$ to $75\Msun$ in steps of $1\Msun$. The testing dataset also contained approximately 2500 templates with intermediate component masses separated from the training set by $0.5\Msun$ each. Subsequently, we produced copies of each signal by shifting the location of their peaks randomly within the final 0.2 seconds to make the CNNs more resilient to time translations. This means that in practice, our algorithm will be applied to the continuous data stream using a 1-second sliding window with offsets of 0.2 seconds. 
	
	We obtained real LIGO data from the LIGO Open Science Center (LOSC) around the first 3 GW events, namely, GW150914, LVT151012, and GW151226. Each event contained 4096 seconds of real data from each detector. We used noise sampled from GW151226 and LVT151012 for training and validation of our model and noise from GW150914 was used for testing. These tests ensure that our method is able to generalize to different noise distributions, also in the presence of transient glitches, since it is well known that the PSD of LIGO is highly non-stationary, varying widely with time. Therefore, if \texttt{Deep Filtering} performs well on these test sets, it would also perform well on data from future time periods, without being re-trained.
	
		\begin{figure}
		\hspace{-.14in}	\includegraphics[width=.48\textwidth]{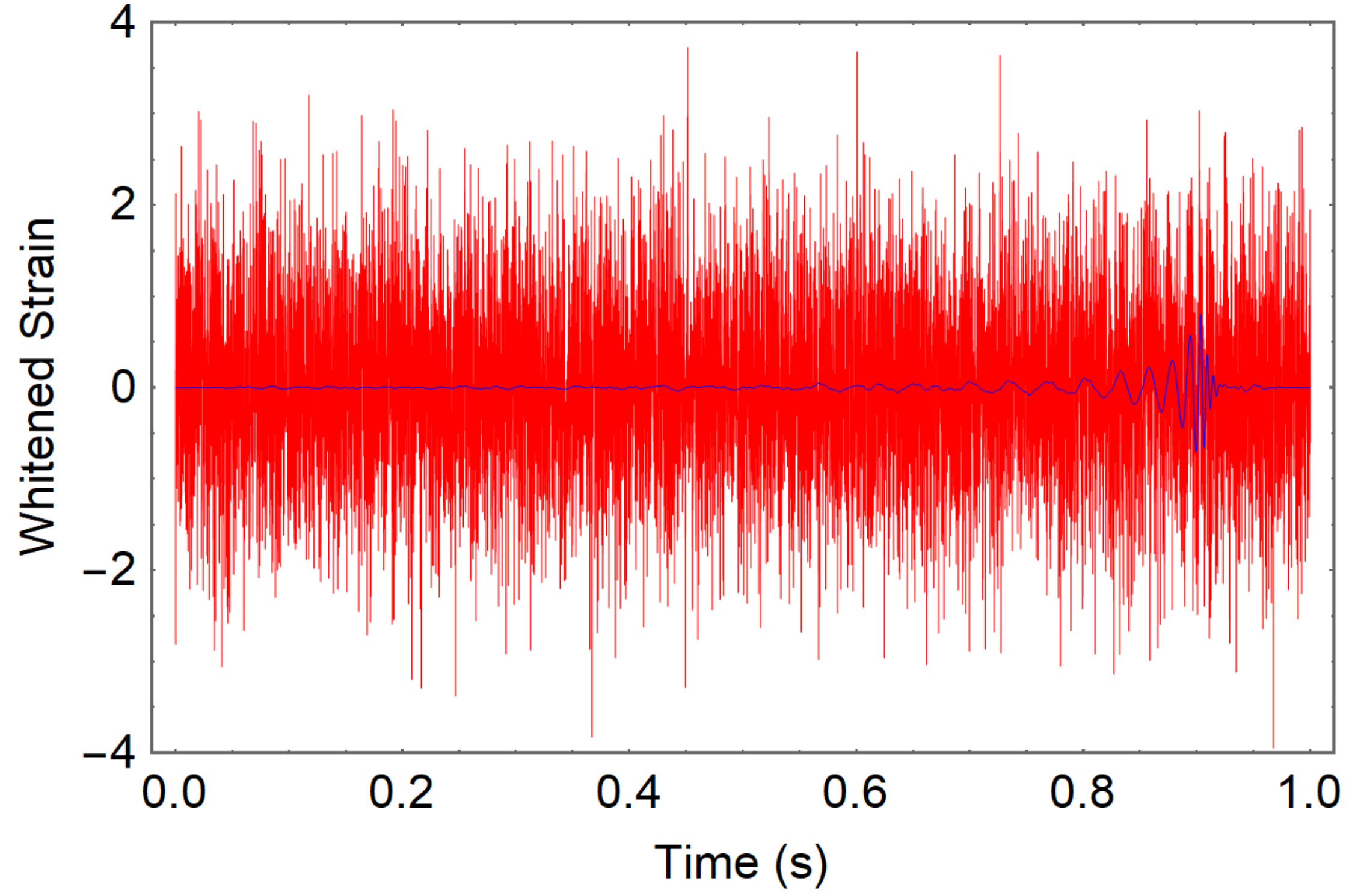}
		
		\caption{\textbf{Sample signal injected into real LIGO noise}. The red time-series is an example of the input to our \texttt{Deep Filtering} algorithm. It contains a hidden BBH GW signal (blue) from our test set which was superimposed in real LIGO noise from the test set and whitened. For this injection, the optimal matched-filter SNR = 7.5 (peak power of this signal is 0.65 times the power of background noise). The component masses of the merging BHs are $57\Msun$ and $33\Msun$. The presence of this signal was detected directly from the (red) time-series input with over 99\% sensitivity and the source's parameters were estimated with a mean relative error less than 10\%.}
		\label{fig:signal}
	\end{figure}

	Next, we superimposed different realizations of noise randomly sampled from the training set of real LIGO noise from the two events GW151226 and LVT151012 and injected signals over multiple iterations, thus amplifying the size of the training datasets. The power of the noise was adjusted according to the desired optimal matched-filter Signal-to-Noise Ratio (SNR~\cite{saton}) for each training round. The inputs were then whitened with the average PSD of the real noise measured at that time-period. We also scaled and mixed different samples of LIGO noise together to artificially produce more training data and various levels of Gaussian noise was also added to augment the training process. However, the testing results were measured using only pure LIGO noise not used in training with true GW signals or with signals injected from the unaltered test sets (see Fig.~\ref{fig:signal}).

	\begin{figure}
		\centering
		\hspace{-.125in}\includegraphics[width=0.5\textwidth]{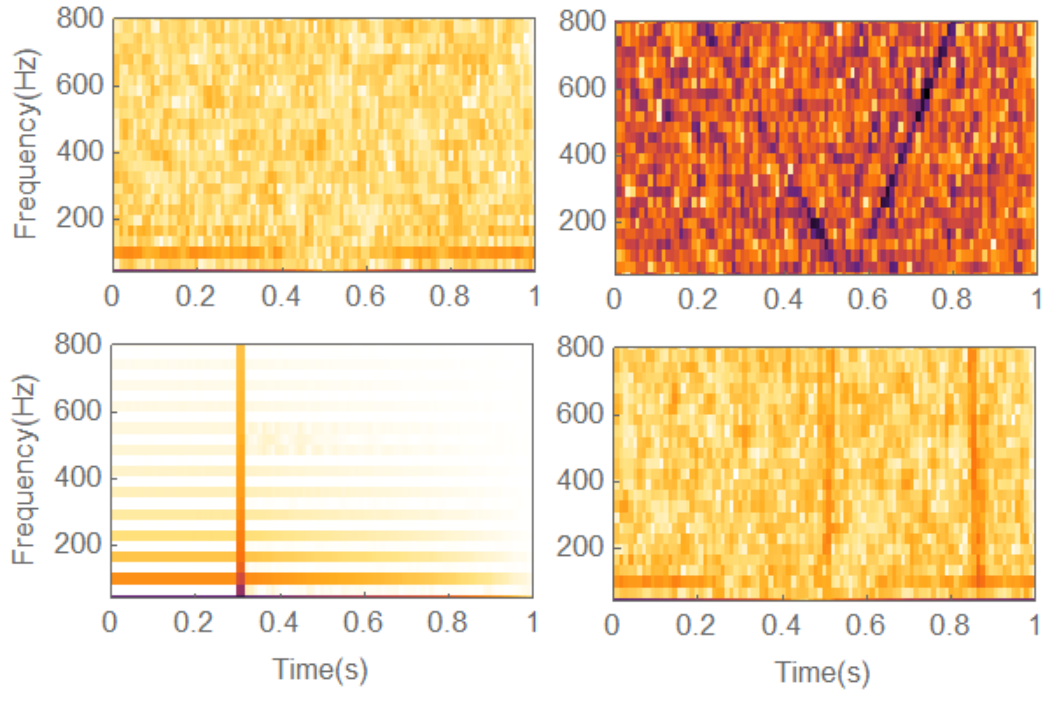}
		\caption{\textbf{Spectrograms of real LIGO noise test samples}. We used signals injected into real data from the LIGO detectors in this article, ensuring that the training and testing sets did not contain noise from the same events. These are some random examples of real glitches that were present in our test set of LIGO noise. The \texttt{Deep Filtering} method takes the 1D strain directly as input and is able to correctly classify glitches as noise and detect true GW signals as well as simulated GW signals injected into these highly non-stationary non-Gaussian data streams, with similar sensitivity compared to matched-filtering.}
		\label{fig:RealGlitches}
	\end{figure}
	
	
	\begin{figure}
		\centering
		\includegraphics[width=0.37\textwidth]{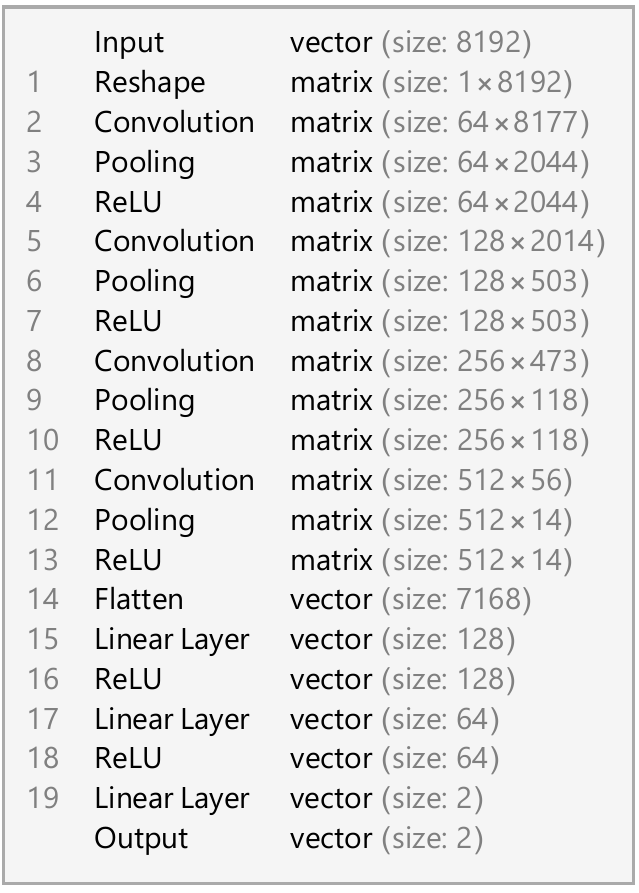}
		\caption{\textbf{Architecture of deep convolutional neural network}. This is the dilated 1D CNN used as the predictor which outputs the component masses of the BBH system. The classifier has the same architecture, except for a softmax layer added at the end which outputs the probability for the presence of a GW signal. The input is a time-series vector of length 8192 corresponding to 1s of data sampled at 8192Hz. The classifier is applied separately to continuous data streams from each detector using a sliding window. If the classifier detects a signal in coincidence across multiple detectors, then the inputs are fed to the predictor which estimates the parameters of the GW source.}
		\label{fig:Design}
	\end{figure}
	
	We used similar hyperparameters to our original CNNs~\cite{DeepFiltering} with a slightly deeper architecture. There were 4 convolution layers with the filter sizes to 64, 128, 256, and 512 respectively and 2 fully connected layers with sizes 128 and 64. The standard ReLU activation function, max$(0,x)$, was used throughout as the non-linearity between layers. We used kernel sizes of 16, 16, 16, and 32 for the convolutional layers and 4 for all the (max) pooling layers. Stride was chosen to be 1 for all the convolution layers and 4 for all the pooling layers. We observed that using dilations~\cite{dilatedCNN} of 1, 2, 2, and 2 in the corresponding convolution layers improved the performance. The final layout of our predictor CNN is shown in Fig.~\ref{fig:Design}.
	
	
	We had originally optimized this CNN architecture to deal with only Gaussian noise having a flat PSD. However, we later found that this model also obtained the best performance with noise having the colored PSD of LIGO, among all the models we tested. This indicates that our architecture is robust to a wide range of noise distributions. Furthermore, pre-training the CNNs on Gaussian noise (transfer learning) before fine-tuning on the limited amount of real noise prevented over-fitting, i.e., memorizing only the training data without generalizing to new inputs. We used the Wolfram Language neural network functionality, based on the open-source MXNet framework~\cite{MXNet}, that uses the cuDNN library~\cite{cuDNN} for accelerating the training with NVIDIA GPUs. The learning algorithm was again set to ADAM~\cite{ADAM} and other details were the same as before~\cite{DeepFiltering}.
	
	For training, we used the curriculum learning strategy in our first article~\cite{DeepFiltering} to improve the performance and reduce training times of the CNNs while retaining performance at very high SNR. By starting off training inputs having high SNR ($\ge100$) and then gradually increasing the noise in each subsequent training session until a final SNR distributed in the range 4 to 15, we found that the performance of prediction can be quickly maximized for low SNR while retaining performance at high SNR. We first trained the predictor on the datasets labeled with the BBH masses and then copied the weights of this network to initialize the classifier and then trained it on datasets having 90\% pure random noise inputs, after adding a softmax layer. This transfer learning procedure, similar to multi-task learning, decreases the training time for the classifier and improves its sensitivity.
	
	\section{Results} 
	\label{result}

	\begin{figure}
		
	\hspace{-.14in}	\includegraphics[width=.485\textwidth]{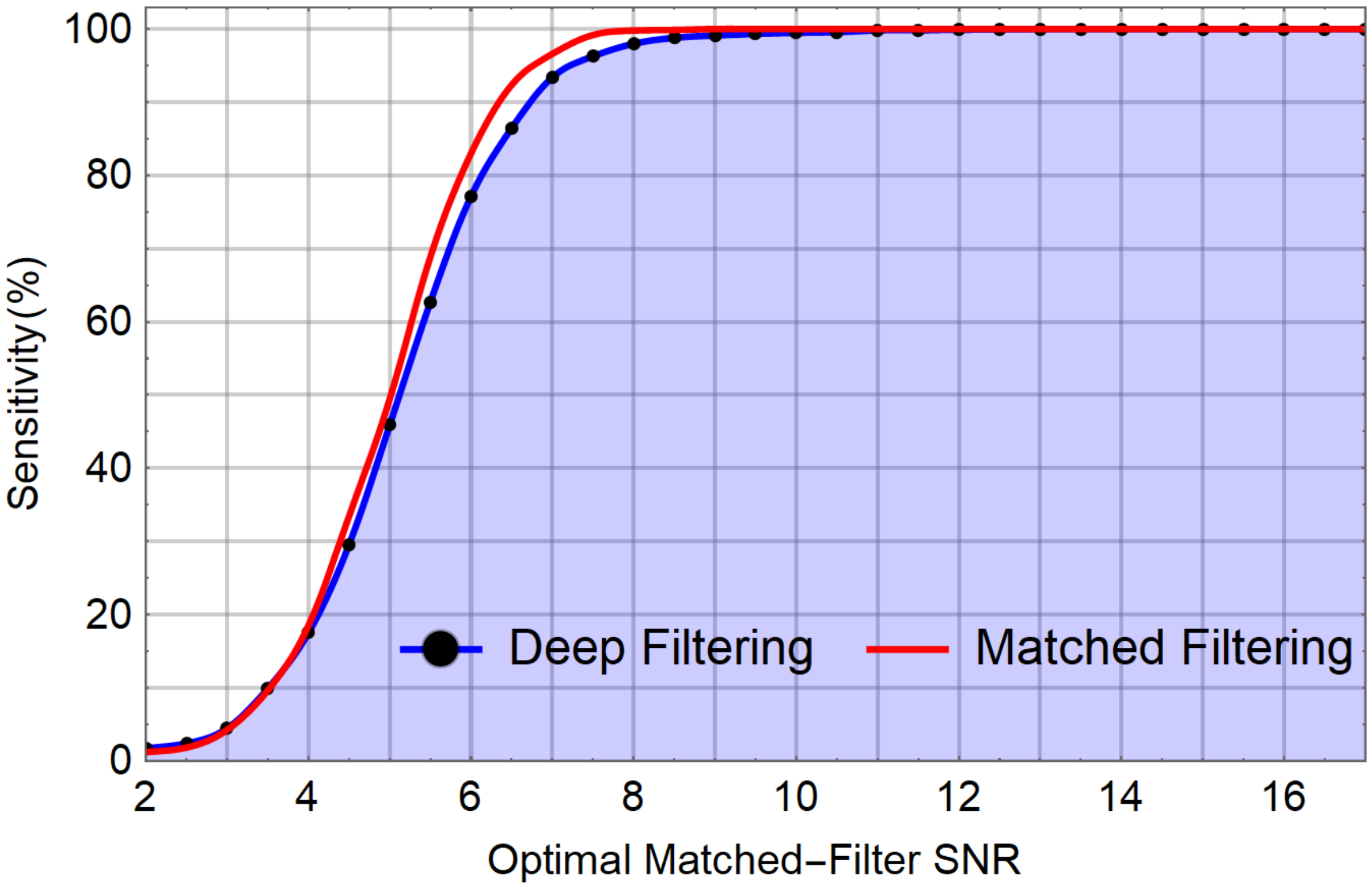}
		
		\caption{\textbf{Sensitivity of detection with real LIGO noise.}
			The curve shows the sensitivity of detecting GW signals injected in real LIGO noise (from LOSC) using Deep Filtering and matched filtering with the same template bank used for training. Note that the SNR is on average proportional to $10\pm1.5$ times the ratio of the amplitude of the signal to the standard deviation of the noise for our test set. This implies that we are capable of detecting signals significantly weaker than the background noise.
		}
		\label{fig:Sensitivity}
	\end{figure}
	
	The sensitivity (probability of detecting a true signal) of the classifier as a function of SNR is shown in Fig.~\ref{fig:Sensitivity}. We achieved 100\% sensitivity when SNR is greater than 10. The false alarm rate was tuned to be less than 1\%, i.e., 1 per 100 seconds of noise in our test set was classified as signals. Given independent noise from multiple detectors, this implies our 2-detector false alarm rate would be less than 0.01\%, when the classifier is applied independently to each detector and coincidence is enforced. Although the false alarm rate can be further decreased by tuning the fraction of noise used for training or by checking that the predicted parameters are consistent, this may not be necessary since running matched-filtering pipelines with a few templates close to our predicted parameters can quickly eliminate these false alarms.
	
	\begin{figure}
		
		\hspace{-.15in}	\includegraphics[width=.48\textwidth]{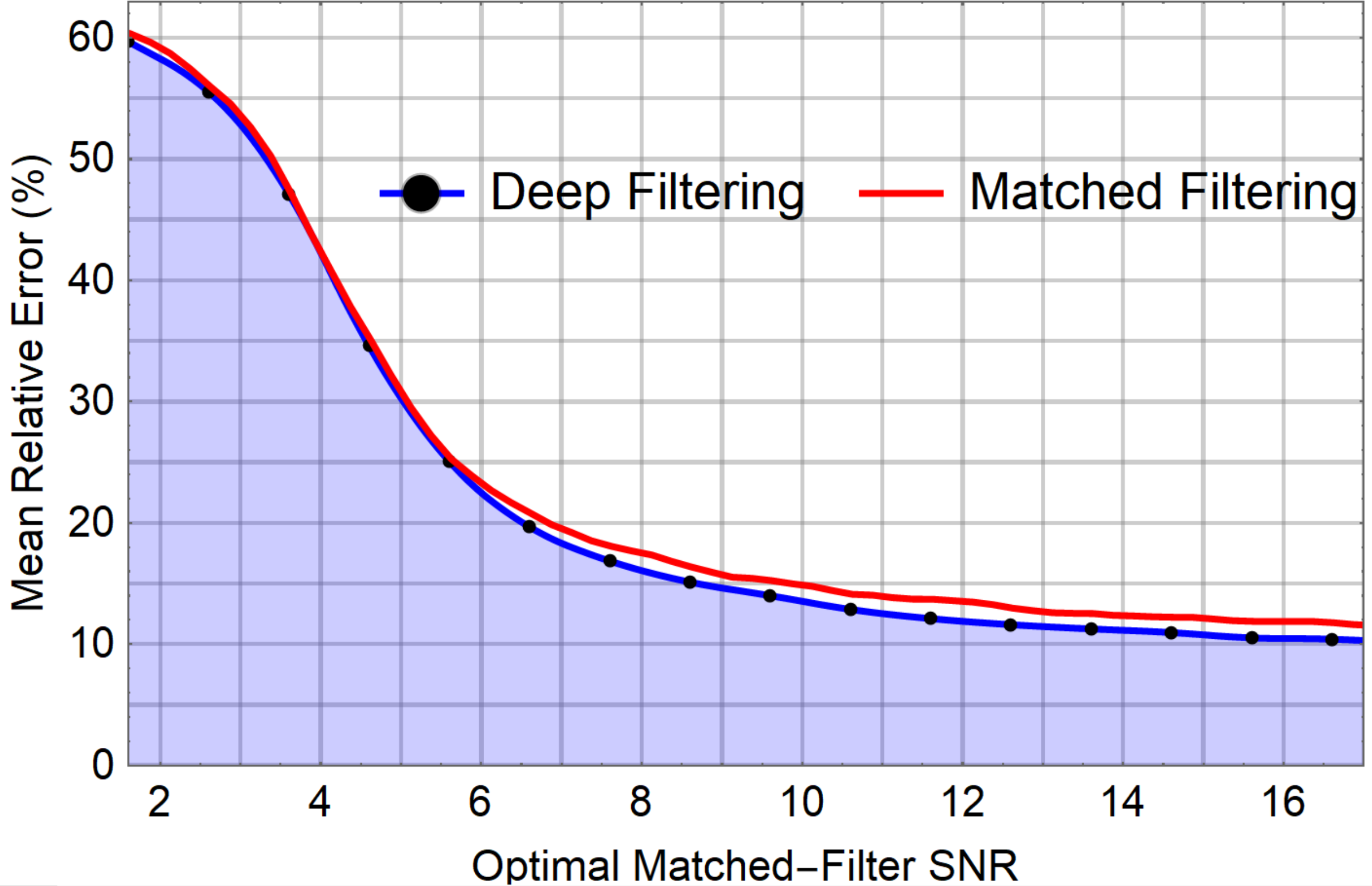}
		
		\caption{\textbf{Error in parameter estimation with real LIGO noise}. This shows the mean percentage absolute error of estimating masses on our testing signals, at each SNR, injected in real LIGO noise compared to matched filtering with the same template bank that was used for training. While the mean error of matched-filtering, with the same template bank used for training, is always greater than 11\% at every SNR we can see that the Deep Filtering method is able to interpolate to test set signals with intermediate parameter values.
		}
		\label{fig:Error}
	\end{figure}
	
	Our predictor was able to successfully measure the component masses given noisy GW signals, that were not used for training, with an error lower than the spacing between templates for optimal matched-filter SNR $\ge15.0$. The variation in relative error against SNR is shown in Fig.~\ref{fig:Error}. We observed that the errors follow a Gaussian distribution for each region of the parameter space for SNR greater than 10. For high SNR, our predictor achieved mean relative error less than 10\%, whereas matched-filtering with the same template bank always has error greater than 10\%. This implies that Deep Filtering is capable of interpolating between templates seen in the training data.
	
	Although, we trained only on simulated quasi-circular non-spinning GW injections, we applied \texttt{Deep Filtering} to the LIGO data streams containing a true GW signal, GW150914, using a sliding window of 1s width with offsets of 0.2s through the data around each event from each detector. This signal was correctly identified by the classifier at the true position in time and each of the predicted component masses were within the published error bars~\cite{DI:2016}. There were zero false alarms after enforcing the constraint that the detection should be made simultaneously in multiple detectors. This shows that deep learning is able to generalize to real GW signals after being trained only with simulated GW templates injected into LIGO noise from other events with different PSDs. A demo showing the application of \texttt{Deep Filtering} to GW150914 can be found here: \href{http://tiny.cc/CNN}{tiny.cc/CNN}.
	
	The data from the first LIGO event, that was used for testing, contained a large number of non-Gaussian transient noise called glitches. Some of these can be seen in Fig.~\ref{fig:RealGlitches}. Therefore, our results demonstrate that the \texttt{Deep Filtering} method can automatically recognize these glitches and classify them as noise. This suggests that by adding additional neurons for each ``glitch'' class, Deep Filtering could serve as an alternative to glitch classification algorithms based on two-dimensional CNNs applied to spectrograms of LIGO~\cite{GravitySpy,DeepTransferLearning} or machine learning methods based on manually engineered features~\cite{DBNN,jade1:2016,jade:2015CQGra}.
	
\begin{figure}
		\centering
		\hspace{-.0651in}\includegraphics[width=0.49\textwidth]{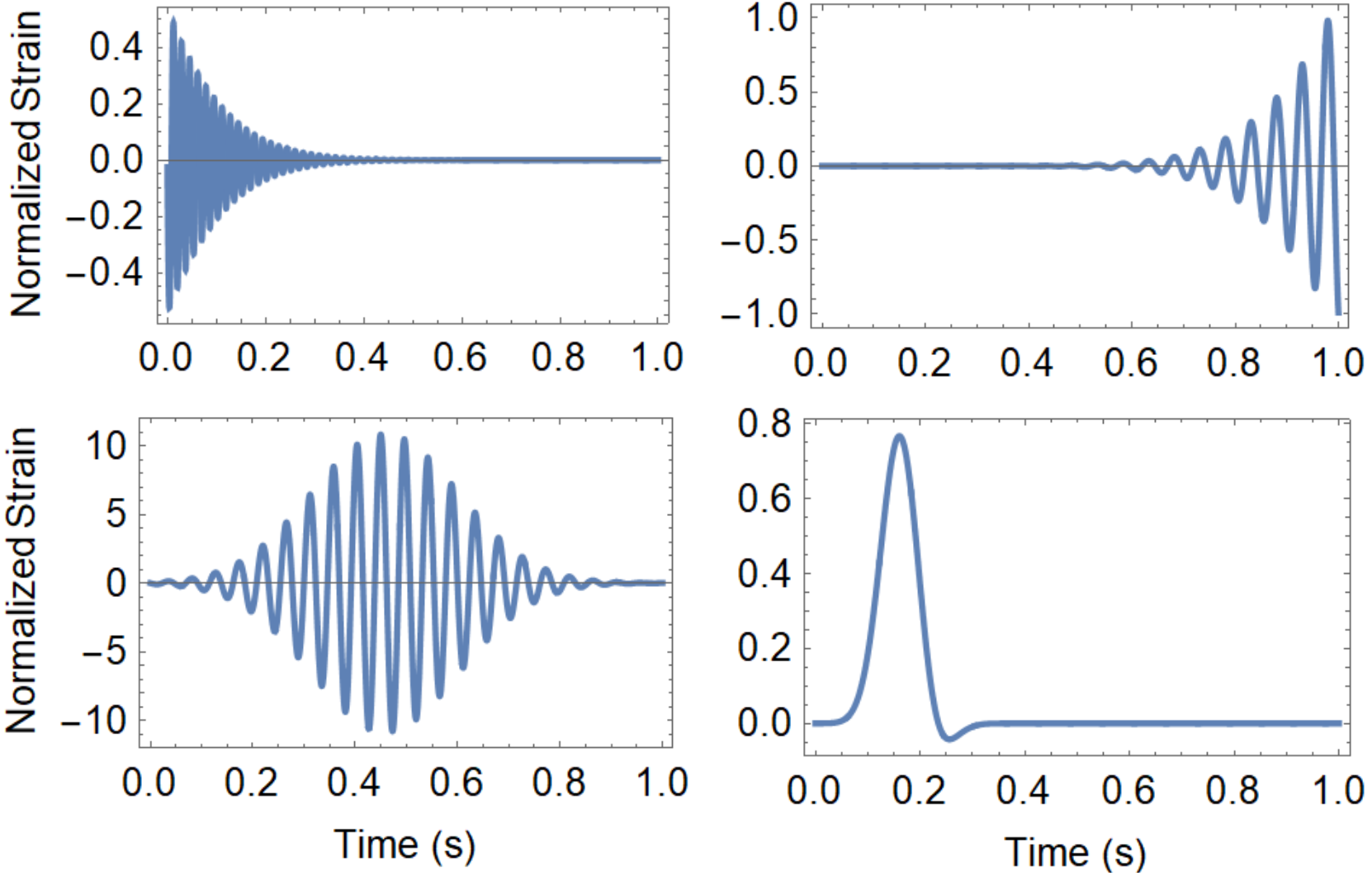}
		\caption{\textbf{Examples of sine-Gaussian glitches}. These are some samples of simulated sine-Gaussian glitches from our test set. We found that our classifier was able to correctly differentiate GW signals from these glitches and classify them as noise when they were injected into real LIGO data streams. This suggests that \texttt{Deep Filtering} can be extended to create a unified pipeline for glitch classification along with signal detection and parameter estimation.}
		\label{fig:Glitch}
	\end{figure}

	Furthermore, we conducted some experiments to show the resilience of \texttt{Deep Filtering} to transient disturbances, with a simulated set of sine-Gaussian glitches, which cover a broad range of morphologies found in real LIGO glitches, following~\cite{jade1:2016} (see Fig.~\ref{fig:Glitch} for some examples). We ensured that a different set of frequencies, amplitudes, peak positions, and widths were used for training and testing. We then injected some of these glitches into the training process and found that the classifier CNN was able to easily distinguish new glitches from true signals, with a false alarm rate less than 1\%. When we applied the standard naive matched-filtering algorithm to the same test set of glitches, approximately 30\% of glitches were classified as signals due to their high SNR. Note that signal consistency tests and coherence across detectors can be enforced to decrease this false alarm rate for both methods.
		
	\begin{figure}
		\centering
		\hspace{-.07in}\includegraphics[width=0.48\textwidth]{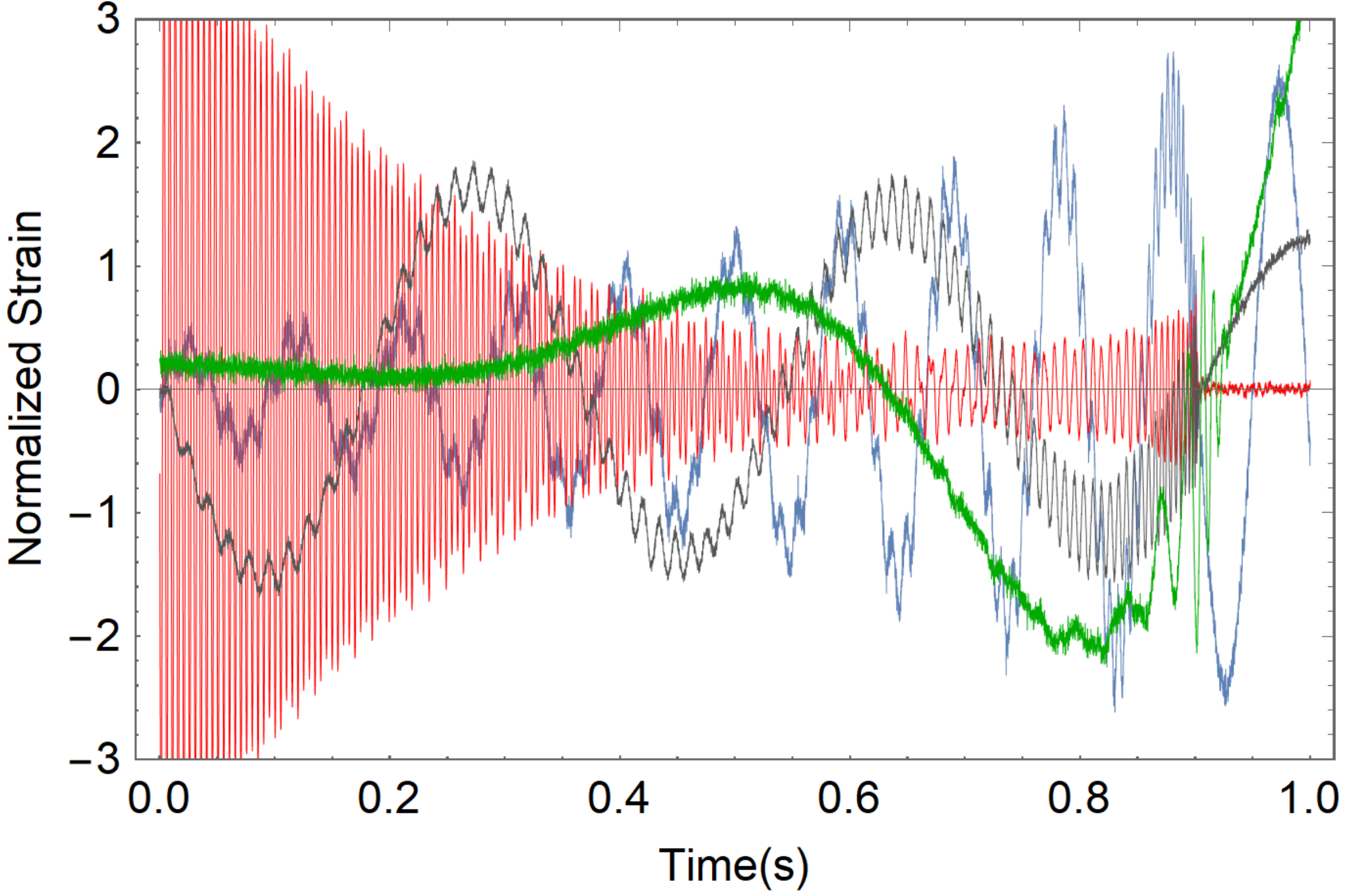}
		\caption{\textbf{Detecting signals contaminated by glitches}. These are some of the signals in our test set injected into real LIGO noise and superimposed with simulated sine-Gaussian glitches from the test set. Each of these inputs were correctly detected as a signal by our classifier. This indicates that \texttt{Deep Filtering} can be used as an automatic trigger generator for GW signals that occur in coincidence with glitches.}
		\label{fig:SignalGlitch}
	\end{figure}
	
	We then tested the performance of \texttt{Deep Filtering}, when a signal happens to occur in coincidence with a glitch, i.e., the signal is superimposed with both a glitch and real LIGO noise. We trained the network by injecting glitches from the training set into the training process and measured the sensitivity of the classifier on the test set signals superimposed with glitches sampled from the test set of glitches. We found that over 80\% of the signals with SNR of 10 were detected, and their parameters estimated with less than 30\% relative error, even after they were superimposed with glitches. These results are very promising, motivating further in-depth investigation, since we may be able to automatically detect GW signals that occur during periods of bad data quality in the detectors using \texttt{Deep Filtering}, whereas currently such periods are often vetoed and left out of the analysis by LIGO pipelines.
	
	Another important experiment that we carried out was to inject waveforms obtained from simulations of eccentric BBH systems, with eccentricities between 0.1 and 0.2 when entering the LIGO band, that we performed using the open-source Einstein Toolkit~\cite{ETL:2012CQGra} as well as waveforms from spin-precessing binaries from the public SXS catalog~\cite{chu:2016CQG}. We found that these signals were detected with the same sensitivity as the original test set of quasi-circular BBH waveforms by our classifier, thus demonstrating its ability to automatically generalize to new classes of GW signals.
	
	Both our CNNs are only 23MB in size each, yet encodes all the relevant information from about 2,500 GW templates (\textasciitilde 300MB) of templates and several GB of noise used to generate the training data. The time-domain matched-filtering algorithm used for comparison required over 2s to analyze 1s inputs on our CPU. The average time taken for evaluating each of our CNNs per second of data is approximately 85 milliseconds and 540 microseconds using a single CPU core and GPU respectively, thus enabling analysis even faster than real-time. While the computational cost of matched-filtering grows exponentially with the number of parameters, the Deep Filtering algorithm requires only a one-time training process, after which the analysis can be performed in constant time. Therefore, we expect the speed-up compared to matched-filtering to further increase by several orders of magnitude when the analysis is extended to larger number of parameters.

	\section{Conclusion}
	\label{conc}
	
	In this article, we have shown for the very first time that CNNs can be used for both \textit{detection} and \textit{parameter estimation} of GW signals in LIGO data. This new paradigm for real-time GW analysis may overcome outstanding challenges regarding the extension of established GW detection algorithms to higher dimensions for targeting a deeper parameter space of astrophysically motivated GW sources. The results of \texttt{Deep Filtering} can be quickly verified via matched-filtering with a small set of templates in the predicted region of parameter space. Therefore, by combining \texttt{Deep Filtering} with well-established GW detection algorithms we may be able to accelerate multimessenger campaigns, pushing the frontiers of GW astrophysics and fully realize its potential for scientific discovery. 
	
	The intrinsic scalability of deep learning can overcome the curse of dimensionality, and take advantage of terabytes of training data. This ability could enable simultaneous GW searches covering millions or billions of templates over the full range of parameter-space that is beyond the reach of existing algorithms. Extending \texttt{Deep Filtering} to predict any number of parameters such as spins, eccentricities, etc., or additional classes of signals or noise, is as simple as adding an additional neuron for each new parameter, or class, to the final layer and training with noisy waveforms with the corresponding labels.  Furthermore, the input dimensions of the CNNs can be enlarged to take time-series inputs from multiple detectors, thus allowing coherent searches and measurements of parameters such as sky locations.
	
	The results presented in this article provide a strong incentive to extend \texttt{Deep Filtering} to cover the parameter space of spin-aligned BBHs on quasi-circular orbits and beyond. This study is underway, and will be described in a subsequent article. In addition to our primary results, we have also presented experiments exhibiting the remarkable resilience of this method for detection in periods of bad data quality, even when GW signals are contaminated with non-Gaussian transients. This motivates including additional classes of real glitches, e.g., from the Gravity Spy project~\cite{GravitySpy}, to the training process to automatically classify and cluster glitches directly from the time-series inputs. Therefore, a single, robust, and efficient data analysis pipeline for GW detectors, based on \texttt{Deep Filtering}, that unifies detection and parameter estimation along with glitch classification and clustering in real-time with very low computational overhead may potentially be built in the near future.
	
	Furthermore, \texttt{Deep Filtering} can be used to accelerate Bayesian parameter estimation methods by constraining the parameter space of new GW detections and provide instant alerts with accurate parameters for EM follow-up campaigns. As deep CNNs excel at image processing, applying the same approach to analyze raw telescope data may also accelerate the subsequent search for transient EM counterparts. Our results also suggest that, given templates of expected signals, \texttt{Deep Filtering} can be used as a generic tool  for efficiently detecting and estimating properties of highly noisy time-domain signals embedded in Gaussian noise or non-stationary non-Gaussian noise, even in the presence of transient disturbances. Therefore, we anticipate that the techniques we have developed for analyzing weak signals hidden in complex noise backgrounds could be useful in many other domains of science and technology.

	\section{Acknowledgements}
	\label{ack}
	This research is part of the Blue Waters sustained-petascale computing project, supported by the National Science Foundation (awards OCI-0725070 and ACI-1238993) and the state of Illinois. Blue Waters is a joint effort of the University of Illinois at Urbana-Champaign and its National Center for Supercomputing Applications. The eccentric numerical relativity simulations used in this article were generated with the open source, community software, the Einstein Toolkit on the Blue Waters petascale supercomputer and XSEDE (TG-PHY160053). We express our gratitude to Gabrielle Allen, Ed Seidel, Roland Haas, Miguel Holgado, Haris Markakis, Zhizhen Zhao and other members of the \href{http://gravity.ncsa.illinois.edu}{NCSA Gravity Group} along with Prannoy Mupparaju for comments and interactions and to the many others who reviewed our manuscript, including Jin Li and Kai Staats. We acknowledge the LIGO/Virgo collaboration, especially the CBC and MLA groups, for their feedback and for use of computational resources. We thank Vlad Kindratenko for granting us access to GPUs and HPC resources in the Innovative Systems Lab at NCSA. We are grateful to Wolfram Research for providing technical assistance and licenses and to NVIDIA for donating several Tesla P100 GPUs, which we used for our analysis. This document has LIGO number P1700379.

	\bibliography{references,references2}

\end{document}